\documentclass[aps,amsmath,showpacs,showkeys]{revtex4}
\usepackage{amssymb,amsfonts,amsthm,indentfirst}
\usepackage{epsfig}
\usepackage{graphicx}
\begin{document}
\title{On the correlation measure of two-electron systems}
\author{Aparna Saha}
\affiliation{Department of Physics, Visva-Bharati University, Santiniketan 731235, India}
\author{B. Talukdar}
\email{binoy123@bsnl.in}
\affiliation{Department of Physics, Visva-Bharati University, Santiniketan 731235, India}
\author{Supriya Chatterjee}
\affiliation{Department of Physics, Bidhannagar College, EB-2, Sector-1, Salt Lake, Kolkata-700064, India}
\begin{abstract}
We make use of a Hylleraas-type wave function to derive an exact analytical model to quantify correlation in two-electron atomic/ionic systems and subsequently employ it to examine the role of inter-electronic repulsion in affecting (i) the bare (uncorrelated) single-particle position- and momentum-space charge distributions and (ii) corresponding Shannon's information entropies. The results presented for the first five members in the helium iso-electronic sequence, on the one hand, correctly demonstrate the effect of correlation on bare charge distributions and, on the other hand, lead us to some important results for the correlated and uncorrelated values of the entropies. These include the limiting behavior of the correlated entropy sum (sum of position- and momentum-space entropies) and geometrical realization for the variation of  information entropies as a function of $Z$. We suggest that, rather than the entropy sum, individual entropies should be regarded as better candidates for the measure of correlation.
\end{abstract}
\maketitle
\section{Introduction}
Our understanding of atomic theory is largely based on the independent-electron model in which the effect of the inter-electronic repulsion, globally referred to as correlation, is disregarded. The correlation effects can, however, have major influence on measureable quantities of atomic systems. The correlation energy, defined by the difference between the exact total energy and Hartree-Fock energy, is traditionally used as a guide \cite{1} for the amount of correlation in a given system. Relatively recently, it has been proposed that Shannon's information entropies \cite{2} provide a useful basis for the measure of electron correlation in atomic systems \cite{3}.
\par For a many-electron atom, Shannon's position-space information entropy is defined by \cite{2} 
\begin{equation}
 S_{\rho}=-\int\rho(\overrightarrow{r})\ln\rho(\overrightarrow{r})d\overrightarrow{r},
\end{equation}
where
\begin{equation}
 \rho(\overrightarrow{r})=\int|\psi(\overrightarrow{r},\,\overrightarrow{r}_2,\,....,\overrightarrow{r}_N)|^2d\overrightarrow{r}_2\,.....d\overrightarrow{r}_N
\end{equation}
with $\psi(\overrightarrow{r},\,\overrightarrow{r}_2,\,....,\overrightarrow{r}_N)$, the normalized wave function of an $N$ electron atom. The charge density $\rho(\overrightarrow{r})$ is also normalized to unity. Correspondingly, the momentum- space entropy is written as
\begin{equation}
 S_{\gamma}=-\int\gamma(\overrightarrow{p})\ln\gamma(\overrightarrow{p})d\overrightarrow{p}.
\end{equation}
The momentum-space single-particle density $\gamma(\overrightarrow{p})$ is similar to that given in (2) but characterized by the normalized momentum-space wave function $\phi(\overrightarrow{p},\,\overrightarrow{p}_2,\,....,\overrightarrow{p}_N)$ obtained by taking Fourier transform of the position-space wave function. The two entropies as introduced through (1) and (3) allowed Bialynicki-Birula and Mycielski \cite{4} to introduce a stronger version of the Heisenberg uncertainty relation which for any 3-dimensional system reads
\begin{equation}
 S_{\rho}+S_{\gamma}\geq3(1+\ln\pi).
\end{equation}
Equation (4), often referred to as BBM inequality, clearly indicates the reciprocity between the representation and momentum spaces such that high values of $S_{\rho}$ are associated with low values of $S_{\gamma}$. For uncorrelated two-electron systems (separable wave functions) the entropy sum is independent of the atomic number $Z$ and has a constant value 6.5665. But when the electron-electron interaction is switched on, the entropy sum appears to depend  on $Z$ \cite{5}. Consequently, the correlated entropy sum $S_{\rho}+S_{\gamma}$ has also been used as measure of correlation in atomic systems.
\par Information entropies of two-electron systems can be expressed in closed analytic form for separable approximation of the wave function. This is, however, not possible for explicitly $r_{12}$ - (inter-electronic separation) dependent wave functions which can account for correlation in the system.The object of the present work is to derive an exact analytic model to quantify correlation in two-electron systems described by wave functions depending explicitly on $r_{12}$. In particular, we shall use our model to study the effect of inter-electronic repulsion on Shannon's information entropies with a view to look for a new measure of correlation. The system of our interest consists of the first five members in the helium iso-electronic sequence.
\par There exist elaborate studies on the problem of determining accurate wave functions for the helium atom \cite{6}. The constructed wave functions often involve very large number of variational parameters. Consequently, it is rather difficult to extend such approaches to deal with other ions in the helium iso-electronic sequence. Moreover, in applicative context, use of these wave functions leads to very large number of computations. In view of this, occasionally, there have been attempts to construct accurate correlated two-electron wave functions involving fewer numbers of parameters. One such attempt is due to Bhattacharyya et al. \cite{7}. An important virtue of their work is that the proposed three-parameter correlated wave function could be used to construct exact analytic expression for the total energy of two-electron systems. But it was noted that the wave function of ref. 7 satisfies cusp conditions \cite{8} only approximately such that the total energy computed for the helium atom is higher than the  experimental value. As a result there have been many attempts to improve on the quality of such few-parameter wave functions. A significant work along this line of investigation is due to Le Sech \cite{9} who constructed a two-parameter wave function which not only satisfies the correlation cusp condition to a high degree of accuracy but also gives much improved values for the energies of two-electron atomic/ionic systems from $H^-$ to $Be^{3+}$. Recently, Chauhan and Harbola \cite{10} reexamined the  optimization procedure for the parameters in the  wave function of Le Sech in order to make it more accurate than that used in ref.9 for the computation of ground-state energies  of two-electron atoms and ions. 
\par The wave function of our interest can be written in the form 
\begin{equation}
 \psi(\overrightarrow{r}_1,\overrightarrow{r}_2,r_{12})=\frac{C_N}{2}e^{-Z(r_1+r_2)}(\cosh(ar_1)+\cosh(ar_2))(1+\frac{\lambda}{2}r_{12}e^{-br_{12}})
\end{equation}
with $Z$, the atomic number of the system and $C_{N}$, the normalization constant of the wave function. Here $a$ and $b$ are variational parameters. For $\lambda=1$ (5) gives the wave function of Le Sech \cite{9} while for $\lambda=0$ and $a=0$ we get the well known separable two-electron wave function \cite{11}. We shall construct closed form expressions for both single-particle coordinate- and momentum-space charge densities and then employ them to examine effect of correlation on individual entropies and entropy sum $S_{\rho}+S_{\gamma}$ by using the parameters of Chauhan and Harbola \cite{10}.
\par We begin Sec. II by noting that instead of using (2) and a similar result written in the momentum space, it is possible to obtain expressions for single-particle charge densities by following an alternative approach. The second viewpoint is particularly well suited to derive results for position- and momentum-space charge densities $\rho(\overrightarrow{r})$ and $\gamma(\overrightarrow{p})$ in the presence of correlation. We demonstrate that, due to the effect of inter-electronic repulsion, the bare or uncorrelated radial charge distribution is pushed apart. On the other hand, the bare momentum distribution is squeezed by the same effect. In Sec. III we present results for the position- and momentum-space information entropies computed on the basis of our expressions for $\rho(\overrightarrow{r})$ and $\gamma(\overrightarrow{p})$. We verify that the uncorrelated entropy sum is a constant which does not depend on the atomic number \cite{5}. As opposed to this, in the presence of correlation the entropy sum becomes $Z$ dependent. However, correlated entropy sum tends to the un-correlated $Z$-independent constant value as we go along the iso-electronic sequence. The position-space entropy $S_{\rho}$ takes up negative values for high-$Z$ atoms but values of momentum-space entropy $S_{\gamma}$ are always positive. We provide a geometrical realization for this fact and thus try to visualize how the interplay between nuclear Coulomb interaction and inter-electronic repulsion determines the variation of $S_{\rho}$ or $S_{\gamma}$ as a function of $Z$. We observe that individual entropies are more strongly affected by the inter-electronic repulsion than the corresponding entropy sum such that the Shannon's information entropy is a very suitable candidate for the measure of correlation. Finally, we make some concluding remarks in Sec. IV.
\section{Single-particle charge densities for two-electron systems}
We shall first work with an uncorrelated or separable wave function for the two-electron system and show that, rather than using (2) or its momentum-space analog, it is possible to follow a different route to construct expressions for charge densities. The well known two-electron separable wave function  is given by \cite{11} 
\begin{equation}
 \psi(\overrightarrow{r},\overrightarrow{r}_2)=ce^{-Z(r+r_2)}
\end{equation}
with $c$ an appropriate normalization constant. From (2) and (6) it is straightforward to show that 
\begin{equation}
 \rho(\overrightarrow{r})=\frac{Z^3}{\pi}e^{-2Zr}
\end{equation}
represents the normalized single-particle charge density of the system.  The result in (7) could also be derived by first constructing a single particle wave function
\begin{equation}
 \psi(\overrightarrow{r})=\frac{8\pi c}{Z^3}e^{-Zr} 
\end{equation}
by integrating (6) over $\overrightarrow{r}_2$ and subsequently normalizing  the wave function in (8) to unity. 
\par It is straightforward to verify that the alternative approach presented above is equally applicable for the momentum-space ($p$-space) separable wave function
\begin{equation}
 \phi(\overrightarrow{p},\overrightarrow{p}_2)=\frac{\tilde{c}}{(p^2+Z^2)(p_2^2+Z^2)}
\end{equation}
obtained by taking the Fourier transform of (6). Here $\tilde{c}$ is the normalization constant of the momentum-space 
wave function. This implies that the momentum-space charge density $\gamma(\overrightarrow{p})$ computed from the wave function $\phi(\overrightarrow{p})(=\int\phi(\overrightarrow{p},\overrightarrow{p}_2)d\overrightarrow{p}_2)$ is exactly equal to that found from the momentum-space analog of (2). 
\par We shall now follow the second viewpoint to derive expressions for charge densities of correlated two-electron systems described by non-separable wave functions. To that end we introduce, for brevity $\overrightarrow{r}_1=\overrightarrow{r}$, and decompose (5) as a sum of two parts
\begin{equation}
 \psi(\overrightarrow{r},\overrightarrow{r}_2,r_{12})=\psi_1(\overrightarrow{r},\overrightarrow{r}_2)+\psi_2(\overrightarrow{r},\overrightarrow{r}_2,r_{12})
\end{equation}
such that
\begin{equation}
\psi_1(\overrightarrow{r},\overrightarrow{r}_2)=\frac{C_N}{2}e^{-Z(r+r_2)}[\cosh(ar)+\cosh(ar_2)]
\end{equation}
and
\begin{equation}
\psi_2(\overrightarrow{r},\overrightarrow{r}_2,r_{12})=\frac{C_N\lambda}{4}e^{-Z(r+r_2)}[\cosh(ar)+\cosh(ar_2)]r_{12}e^{-br_{12}}.
\end{equation}
Equation (11) can easily be integrated over the variable $\overrightarrow{r}_2$ to get
\begin{equation}
\psi_1({\overrightarrow{r}})=2\pi C_Ne^{-Zr}[\frac{1}{(Z-a)^3}+\frac{1}{(Z+a)^3}+\frac{1}{Z^3}(e^{ar}+e^{-ar})].
\end{equation}
It is, however, nontrivial to integrate (12) because $r_2$ occurs here as an entangled variable. This poses an awkward analytical constraint to derive an expression for the single particle charge density by using non-separable wave functions. Fortunately, we can make use of the identity \cite{7}
\begin{equation}
 \frac{e^{-br_{12}}}{r_{12}}=\frac{1}{2\pi^2}\int\frac{e^{i\overrightarrow{q}.(\overrightarrow{r}_1-\overrightarrow{r}_2)}}{b^2+q^2}d\overrightarrow{q}
\end{equation}
to circumvent the difficulty. For example, from (14) we can deduce that
\begin{equation}
 r_{12}e^{-br_{12}}=\frac{1}{\pi^2}\int\frac{(3b^2-q^2)}{(b^2+q^2)^3}e^{i\overrightarrow{q}.\overrightarrow{r}_1}e^{-i\overrightarrow{q}.\overrightarrow{r}_2}d\overrightarrow{q}
\end{equation}
which when substituted in (12) provides a separable representation of $\psi_2(\overrightarrow{r},\overrightarrow{r}_2,r_{12})$ such that
\begin{equation}
 \psi_2(\overrightarrow{r},\overrightarrow{r}_2,r_{12})=\frac{C_N\lambda e^{-Zr}}{4\pi^2}(A(r,r_2)+B(r,r_2))
\end{equation}
with
\begin{equation}
A(r,r_2)=\cosh(ar)\int\frac{3b^2-q^2}{(b^2+q^2)^3}e^{i\overrightarrow{q}.\overrightarrow{r}}e^{Zr_2-i\overrightarrow{q}.\overrightarrow{r}_2}d\overrightarrow{q}
\end{equation}
and
\begin{equation}
B(r,r_2)=\int\frac{3b^2-q^2}{(b^2+q^2)^3}e^{i\overrightarrow{q}.\overrightarrow{r}}\cosh(ar_2)e^{Zr_2-i\overrightarrow{q}.\overrightarrow{r}_2}d\overrightarrow{q}. 
\end{equation}
Equations (17) and (18) can be integrated over the variable $r_2$ by making use of \cite{12}
\begin{equation}
 \int e^{-\gamma \xi+i\overrightarrow{\mu}.\overrightarrow{\xi}}d\overrightarrow{\xi}=\frac{8\pi\gamma}{(\gamma^2+\mu^2)^2}.
\end{equation}
This gives
\begin{equation}
\psi_2(\overrightarrow{r})=\frac{C_N\lambda e^{-Zr}}{\pi}(F(r)+G(r)) 
\end{equation}
with
\begin{equation}
F(r)=2Z\cosh(ar)\int\frac{(3b^2-q^2)e^{i\overrightarrow{q}.\overrightarrow{r}}}{(b^2+q^2)^3(Z^2+q^2)^2}d\overrightarrow{q}
\end{equation}
and
\begin{equation}
G(r)=(Z-a)\int\frac{(3b^2-q^2)e^{i\overrightarrow{q}.\overrightarrow{r}}}{(b^2+q^2)^3((Z-a)^2+q^2)^2}d\overrightarrow{q}+(Z+a)\int\frac{(3b^2-q^2)e^{i\overrightarrow{q}.\overrightarrow{r}}}{(b^2+q^2)^3((Z+a)^2+q^2)^2}d\overrightarrow{q}.
\end{equation}
The angular integrals in (21) and (22) can be evaluated with the help of
\begin{equation}
 \int e^{i\overrightarrow{q}.\overrightarrow{r}}d\Omega_q=\frac{4\pi}{qr}\sin(qr).
\end{equation}
And, finally integration over $q$ yields
\begin{equation}
F(r)=-\frac{\pi^2}{r}\cosh(ar)\frac{\partial^2}{\partial b^2}\frac{\partial}{\partial Z}(\frac{e^{-br}-e^{-Zr}}{Z^2-b^2})
\end{equation}
and
\begin{equation}
G(r)=-\frac{\pi^2}{2r}\frac{\partial^2}{\partial b^2}\frac{\partial}{\partial Z}(\frac{e^{-br}-e^{-(Z-a)r}}{(Z-a)^2-b^2})-\frac{\pi^2}{2r}\frac{\partial^2}{\partial b^2}\frac{\partial}{\partial Z}(\frac{e^{-br}-e^{-(Z+a)r}}{(Z+a)^2-b^2}). 
\end{equation}
As with (9), the results in (13) and (20) together with (24) and (25) give a single-electron wave function corresponding to the non-separable or entangled state (5). Understandably, the present single-electron wave function obtained by eliminating $\overrightarrow{r}_2$ from the two-particle one will involve the effect of correlation. The uncorrelated wave function can be obtained from (13) and (20) using $\lambda=0$ and $a=0$. The integral in (19) in conjunction with
\begin{equation}
 \int\frac{1}{\xi} e^{-\gamma \xi+i\overrightarrow{\mu}.\overrightarrow{\xi}}d\overrightarrow{\xi}=\frac{4\pi}{(\gamma^2+\mu^2)}
\end{equation}
and
\begin{equation}
 \int \xi e^{-\gamma \xi+i\overrightarrow{\mu}.\overrightarrow{\xi}}d\overrightarrow{\xi}=\frac{8\pi(3\gamma^2-\mu^2)}{(\gamma^2+\mu^2)^3}
\end{equation}
can now be used to obtain the momentum-space wave function from our constructed coordinate-space result. Both coordinate- and momentum-space wave functions involve normalization constants. If we fix their values by normalizing the corresponding wave functions to unity, we can write expressions for coordinate- and momentum-space charge densities. In this context it is important to note that in our approach we could study the momentum-space properties of two-electron systems using tabulated integrals only. In the past such studies were found to call for the use of Monte Carlo algorithms to calculate the integrals involved \cite{13}. It is also straightforward  to compute numbers for $S_{\rho}$ and $S_{\gamma}$ with the help of our expressions for charge densities. In this context we observe that, in contrast to the case of correlated two-electron systems, it is relatively uncomplicated to construct closed form analytic expressions for position- and momentum-space charge densities as well as corresponding  information entropies for the one-electron atom \cite{14}. The polynomials occurring in the wave functions of hydrogenic excited states, however, tend to present difficulties to compute results for associated information entropies. One of us \cite{15} demonstrated that the  problem can be circumvented by the simultaneous use of series and product representations of the polynomials \cite{16}.
\par We shall first examine the effect of inter-electronic repulsion on the charge-densities by using the parameters of the wave function (5) as given in ref. 10. In Table 1 we  present numbers for  correlated and uncorrelated charge distributions at $r=0$ (coordinate space) and $p=0$ (momentum space) for the two-electron systems from $H^-$ to $B^{3+}$. For ready reference, we also include in this table the parameters of the wave function. Understandably, the uncorrelated charge distributions are obtained from the corresponding correlated distributions using $\lambda=0$ and $a=0$.
\begin{table}[ht]
\begin{center}
\begin{tabular}{|l|l|l|l|l|}
\hline
Atom & $a$ & $b$ & $\rho(r=0)$ & $\gamma(p=0)$ \\
\hline
$H^-$ & 0.58 & 0.06 & (0.2641) & (2.4051) \\
      &      &      &  0.3183  &  0.8106\\
\hline
$He$ & 0.72 & 0.20 & (2.0164) & (0.1815) \\
      &      &      & 2.5464  &  0.1013\\
\hline
$Li^+$ & 0.87 & 0.36 & (7.4649) & (0.0455) \\
      &      &      & 8.5944  &  0.0300\\
\hline
$Be^{2+}$ & 0.99 & 0.52 & (18.2613) & (0.0172) \\
      &      &      & 20.3718  &  0.0127\\
\hline
$B^{3+}$ & 1.1 & 0.67 & (36.3770) & (0.0095) \\
      &      &      & 39.7887  &  0.0065\\
\hline
\end{tabular}
\caption{Parameters of the wave function (5) and values of the coordinate- and momentum-space charge densities at $r=0$ and $p=0$.}
\end{center}
\end{table}
\begin{figure}[ht]
\begin{center}
\includegraphics[width=7.5cm]{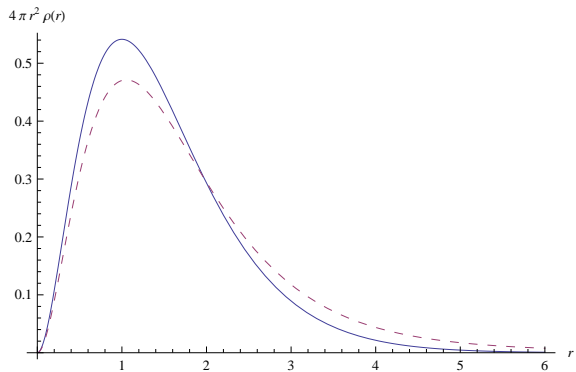}
\caption{(Color online) Coordinate-space charge density $\rho(r)$ of $H^-$ ion as a function of $r$.}.
\end{center}
\end{figure}
\begin{figure}[ht]
\begin{center}
\includegraphics[width=6.5cm]{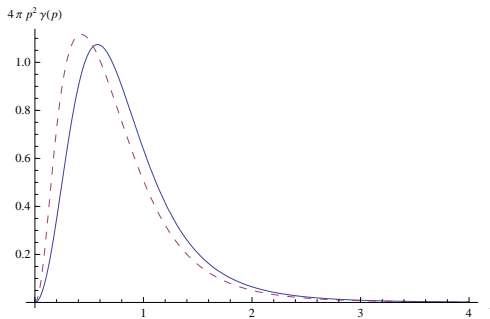}
\caption{(Color online) Momentum-space charge density $\gamma(p)$ of $H^-$ ion as a function of $p$.}
\end{center}
\end{figure}
From Table 1 we see that both parameters of the wave function are monotonically increasing function of the atomic number. The results for coordinate- and momentum-space charge densities at $r=0$ and $p=0$ are presented in columns 4 and 5 of the Table. The numbers for the correlated charge densities are shown in braces. Henceforth we shall use subscript $c$ for the correlated quantities and $0$ for uncorrelated quantities. Looking at the results of the coordinate-space charge densities we see that for all values of $Z$,  $\rho_c(0)< \rho_0(0)$. On the other hand, the results in column 5 show that for all $Z$, $\gamma_c(0)> \gamma_0(0)$. Thus we find that correlation plays an opposite role in modifying the values of the coordinate-space and momentum-space charge densities at $r=0$ and $p=0$. It, therefore, remains an interesting curiosity to examine the overall effect of correlation on the coordinate- and momentum-space charge densities. In order to see this  we display in FIG.1 the results for $4\pi r^2\rho(r)$ as a function of $r$ for $Z=1$. The solid curve denotes the variation of the uncorrelated charge density as a function of $r$ while the dashed curve represents similar variation for the correlated charge density. From this figure it is clear that inter-electronic repulsion pushes the charge distribution away from the origin such that the values of $\rho_c(r)$ are bigger than those of $\rho_0(r)$ for $r>2$. We have verified that the deviation between the solid and dashed curves  diminishes as we go to high $Z$ atoms. This is physically realizable because effect of correlation tends to play less dominant role as we move along the iso-electronic sequence. To examine the role of correlation in modifying the momentum-space  charge density we plot in FIG. 2 $4\pi p^2\gamma(p)$ as a function of $p$. As before, the solid and dotted curves refer to variation of uncorrelated and correlated distributions. We shall follow this convention throughout. Rather than being flattened, the bare charge distribution in this case is squeezed due to inter-electronic repulsion such that the values of the correlated charge density are bigger than those of the uncorrelated density for small $p$ values. For atoms in the helium isoelectronic sequence one can find in the literature a large number of systematic approaches to obtain accurate results for single-particle charge densities in the position space. Contrarily,  there is a lack of similar works on their momentum-space counterpart. For example, a set of benchmark results for the coordinate-space densities was reported by Koga et al \cite{17} even in early 1990's while such a simple analytic model for computing accurate results for the momentum-space densities was reported after about a couple of decades \cite{18}. It is of interest to  note that these detailed studies predicted  effects  of correlation on the bare charge densities, which are in agreement with our findings.
\section{Shannon's position- and momentum-space information entropies}
It is straightforward to use our constructed expressions for $\rho_0(\overrightarrow{r})$, $\rho_c(\overrightarrow{r})$, $\gamma_0(\overrightarrow{p})$ and $\gamma_c(\overrightarrow{p})$ to compute numbers for the corresponding Shannon's position- and momentum- space information entropies $S_{\rho 0}(r)$, $S_{\rho c}(r)$, $S_{\gamma 0}(p)$ and $S_{\gamma c}(p)$. Table 2 gives the results for these entropies for all atoms and ions considered in Table 1. 
\begin{table}[ht]
\begin{center}
\begin{tabular}{|l|l|l|l|l|l|l|}
\hline
Atom & $S_{\rho 0}$ & $S_{\gamma 0}$ &$S_{\rho 0}+S_{\gamma 0}$ & $S_{\rho c}$ & $S_{\gamma c}$&$S_{\rho c}+S_{\gamma c}$ \\
\hline
$H^-$ & 4.1447 & 2.4219 & 6.5666 & 4.6362 & 2.1255 & 6.7617 \\    
\hline
$He$ & 2.0653 & 4.5013 & 6.5666 & 2.4494 & 4.2535 & 6.7029 \\      
\hline
$Li^+$ & 0.8489 & 5.7177 & 6.5666 & 1.1234 & 5.5282 & 6.6516 \\      
\hline
$Be^{2+}$ & -0.0142 & 6.5807 & 6.5665 & 0.1956 & 6.4158 & 6.6114 \\      
\hline
$B^{3+}$ & -0.6836 & 7.2501 & 6.5665 & -0.5123 & 7.0955 & 6.5832 \\      
\hline
\end{tabular}
\caption{Position- and momentum-space information entropies for members of the helium iso-electronic sequence.}
\end{center}
\end{table}  
Looking closely into these numbers we see that the position-space entropies $S_{\rho 0}$ and $S_{\rho c}$ are decreasing function of $Z$ while $S_{\gamma 0}$ and $S_{\gamma c}$ are increasing function of $Z$. Recently, while examining the role of statistical correlation in the N-particle Moshinsky model, Peng and Ho \cite{19} observed a similar behavior of Shannon information entropy in different phase spaces. In two interesting publications Koscik and Okopinska \cite{20} studied the behavior of von Neumann and linear entropies in the Moshinsky model and in helium-like atoms. It is rather interesting to note that the response of these entropies to inter-electronic repulsion appears to  agree with  that exhibited by the position-space Shannon entropies in Table 2. As for the effect of inter-electronic repulsion on the individual entropy we note that correlation plays an opposite role in modifying the values of $S_{\rho 0}$ and $S_{\gamma 0}$. For example, effect of correlation always increases the value of position-space entropies. On the other hand, values of the momentum-space entropies are reduced by inter-electronic repulsion. From the results in column 4, we see that the entropy sum $S_{\rho 0}+S_{\gamma 0}$ for the separable wave function is independent of the atomic number $Z$. This was shown analytically in \cite{5}. From the numbers in column 7 we see that the effect of inter-electronic repulsion increases the entropy sum by about 3 percent for $H^-$ while for $B^{3+}$ a similar increase in the entropy sum is 0.25 percent only. In fact, the correlation contribution to the entropy sum decreases monotonically as we go along the iso-electronic sequence such that for high $Z$ values the correlated entropy sum tends to coalesce with the uncorrelated entropy sum. It is of interest to note that, rather than the entropy sum, the individual entropies $S_{\rho 0}$ and $S_{\gamma 0}$ are more significantly affected by inter-electronic repulsion. For example, $S_{\rho c}$ of $H^-$ is greater than the corresponding result for $S_{\rho 0}$ by about 11.9 percent. For the momentum-space entropy of $H^-$ we note that$S_{\gamma 0}>S_{\gamma c}$ by approximately 12.2 percent. In this context it may be interesting to note that for helium the correlation correction in $S_{\rho c}$ is about ten times of what one needs to obtain the experimental binding energy of $He$ from the corresponding Hartree-Fock energy. Thus we would venture to suggest that, rather than the entropy sum or binding energy of two-electron systems, the individual Shannon's entropies should be regarded as better candidates for the measure of correlation.
\par We shall now make use of the plots of entropy densities 
\begin{equation}
S_{\rho}(r)=-4\pi r^2\rho(r)\ln\rho(r) 
\end{equation}
and
\begin{equation}
S_{\gamma}(p)=-4\pi p^2\gamma(p)\ln\gamma(p)  
\end{equation}
as a function of $r$ and $p$ respectively to provide a physical feeling for why the values of $S_{\rho 0}$ and $S_{\rho c}$ become negative for high value of $Z$ while the numbers for $S_{\gamma 0}$ and $S_{\gamma c}$ always remain positive. To that end we first remember that contribution to entropies as found by integrating (28) and (29) comes from both core and valence regions in the system. In the core region, the effect of nuclear charge dominates and tends to localize the charge distribution. On the other hand, the inter-electronic repulsion in the valence region causes the density distribution to be delocalized. The interplay between localization and delocalization is responsible for change in sign in the values of $S_{\rho}$. 
\begin{figure}[ht]
\begin{center}
\includegraphics[width=6.5cm]{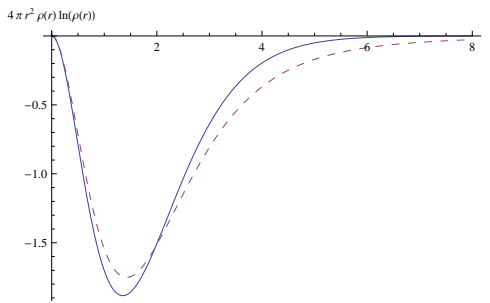}
\caption{(Color online) Entropy densities $-S_{\rho 0}(r)$ and $-S_{\rho c}(r)$ of $H^-$ ion as a function of $r$.}
\end{center}
\end{figure}
We display in FIG. 3 the position-space uncorrelated and correlated entropy densities $-S_{\rho 0}(r)$ and $-S_{\rho c}(r)$ for $H^-$ as a function of $r$. We shall follow this convention for other atoms also. The solid and dotted curves denote the variation of $-S_{\rho 0}(r)$ and $-S_{\rho c}(r)$ respectively. From the curves in this figure it is evident that the results for  entropy densities  are positive for all values of $r$. Physically, this implies that contribution to entropies comes predominantly from inter-electronic repulsion and the core contribution is really insignificant. This is understandable because $H^-$ is a highly delocalized system. From the areas of the solid and dotted curves below the axis it can be shown that $S_{\rho c} > S_{\rho 0}$. This observation is consistent with the results for $H^-$ in Table 2. The deviation between solid and dotted curves increases at large radial distances where the effect of correlation becomes relatively more important. The curves in this figure are not discernible from those reported by Sagar and Guevara \cite{21}.
\begin{figure}[ht]
\begin{center}
\includegraphics[width=6.5cm]{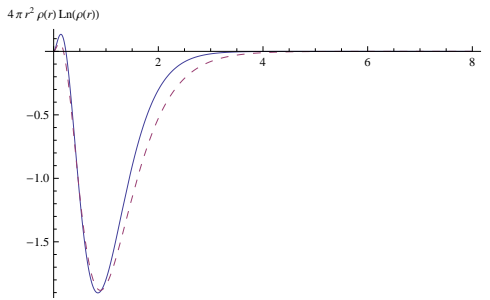}
\caption{(Color online) Entropy densities $-S_{\rho 0}(r)$ and $-S_{\rho c}(r)$ of $He$ ion as a function of $r$.}
\end{center}
\end{figure}
\par In FIG.4 we present similar results for entropy densities for $He$. From this figure we see that both solid and dotted curves show small positive peaks near $r=0$, the peak of the solid curve being little more pronounced. The positive peaks, however small, arise due to localization of the charge density induced by the electron-nucleus interaction. Here the uncorrelated and correlated entropies result from the cancellation between areas of the curves above and below the $r$ -  axis. An important point to note here is that the deviation  between the solid and dotted curves has reduced considerably when compared with the deviation between the corresponding curves of $H^-$. As a result $(S_{\rho c}(r)-S_{\rho 0}(r))_{He}<(S_{\rho c}(r)-S_{\rho 0}(r))_{H^-}$ for all $r$. This is physically realizable because the effect of correlation diminishes as we move to high $Z$ atoms. 
\par The positive peaks in the curves arising due to interaction of the electrons with the nuclear charge  become more and more pronounced for high $Z$ atoms. To visualize this we portray in FIG.5 the entropy densities for $B^{3+}$ as a function of $r$. Here we see that the positive peaks are really very much pronounced. 
\begin{figure}[ht]
\begin{center}
\includegraphics[width=6.5cm]{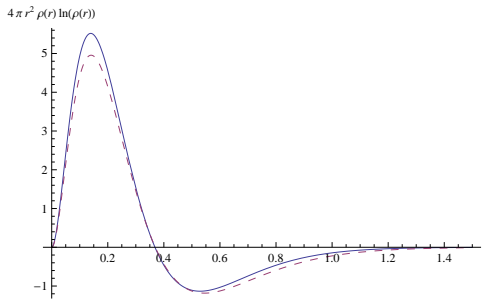}
\caption{(Color online) Entropy densities $-S_{\rho 0}(r)$ and $-S_{\rho c}(r)$ of $B^{3+}$ ion as a function of $r$.}
\end{center}
\end{figure}
The entropies $S_{\rho 0}(r)$ and $S_{\rho c}(r)$ for the triply ionized boron are obtained by the algebraic sum of areas of the solid and dotted curves above and below the $r$- axis. Since the areas of the curves above the $r$ axis are bigger than those below the $r$ axis, the values of both uncorrelated and correlated entropies are negative. As expected due to the effect of correlation the number for $S_{\rho c}$ is greater than that for $S_{\rho 0}$. Here deviation between the solid and dotted curves is appreciable near $r=0.2$ only since the electron-nuclear interaction is quite large for $Z=5$. On the other hand, because of weaker correlation effect, the solid and dotted curves exhibit very little deviation for large $r$ values.
\par In the following we present a similar analysis for numbers of the momentum-space entropies using the plots of appropriate entropy density functions. In FIG.6 we show $-S_{\gamma 0}(p)$ and $-S_{\gamma c}(p)$ for $H^-$ as a function of $p$. As before solid and the dotted curves give the variation of uncorrelated and correlated results.
\begin{figure}[ht]
\begin{center}
\includegraphics[width=6.5cm]{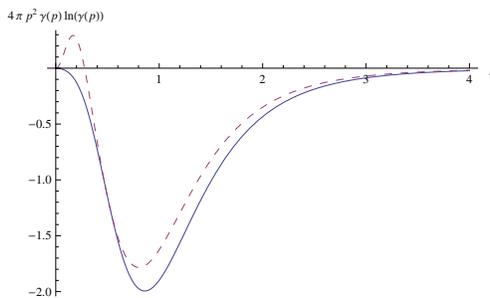}
\caption{(Color online) Momentum-space entropy density $-S_{\gamma 0}(p)$ and $-S_{\gamma c}(p)$ for $H^-$ versus $p$.}
\end{center}
\end{figure}
In close analogy with the solid curve in FIG.3, the curve for the uncorrelated momentum entropy density is negative for all values of $p$. But for  small values of $p$ the dotted curve exhibits a small positive peak then takes up negative values. The negative part of the dotted curve appears to be slightly squeezed compared to the solid curve. Clearly, appearance of the positive peak and observed squeezing account for why the effects of electron-nucleus interaction and of correlation reduce the value of the bare momentum-space entropy. Since the effect of electron-electron interaction gradually diminishes and electron-nucleus interaction increases as we go to high $Z$ atoms, the height of the positive peak in the dotted curve is likely to be modified at large $Z$ values. To see this we display in FIG.7 the momentum-space entropy density of $He$ for which the electron-nucleus interaction is little stronger than the corresponding interaction in $H^-$.
\begin{figure}[ht]
\begin{center}
\includegraphics[width=6.5cm]{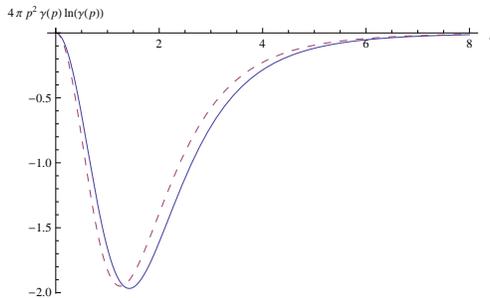}
\caption{(Color online) Momentum-space entropy density $-S_{\gamma 0}(p)$ and $-S_{\gamma c}(p)$ for $He$ versus $p$.}
\end{center}
\end{figure}
\begin{figure}[ht]
\begin{center}
\includegraphics[width=6.5cm]{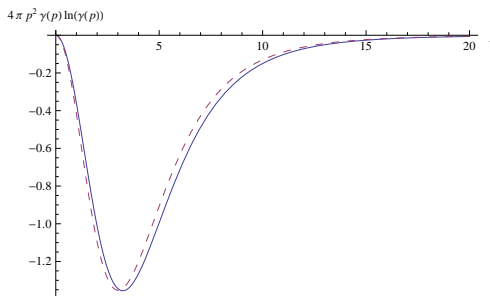}
\caption{(Color online) Momentum-space entropy density $-S_{\gamma 0}(p)$ and $-S_{\gamma c}(p)$ for $Be^{3+}$ versus $p$.}
\end{center}
\end{figure}
From this figure we see that the positive peak in the dotted curve has disappeared due to increase in the strength of the electron-nucleus interaction. In the context of position-space entropy densities we found that the electron-nucleus interaction increases the height of peaks near $r=0$. Here we observe the opposite. In contrast to the plot in FIG. 6, the dotted curve here does not have any peak. This is, however, quite expected since we are now in a reciprocal space. The nature of the solid and dotted curves as appears in FIG.7 does not change for higher atoms in the helium iso-electronic sequence.
This can be seen clearly from the curves in FIG.8 giving the plots of momentum-space entropy densities for $Be^{3+}$.  However, here the area of the solid or dotted curve below the axis is much bigger than those in FIG.7 such that the results for both uncorrelated and correlated entropies of $Be^{3+}$ are greater than the corresponding values for $He$ or $H^-$. Here the deviation between the solid and dotted curves is not appreciable. Consequently, the correlated entropy value differs from the uncorrelated result in Table 3  by about 2 percent only.
\section{Concluding remarks}
It is widely believed that there are distinct advantages to viewing problems of physics within the framework of analytical models, since many physical effects are then readily expressed and evaluated. But in atomic and molecular physics, hardly there are problems that can be solved by using analytical tools. For example, one cannot use realistic atomic wave functions to express single-particle charge densities in closed analytic form for even two-electron systems. In the present paper we work with a highly accurate Hylleraas-type two-electron wave function and derive an analytical model to study the effect of correlation on the single-particle position- and momentum-space charge densities, as well as corresponding Shannon information entropies. With special attention to the first five members of the helium iso-electronic sequence we verify that inter-electronic repulsion flattens the position-space charge density and squeezes the momentum-space charge density. As expected, the effect of correlation is more pronounced for low-$Z$ ions. 
\par The expressions for single-particle charge densities are used to compute numbers for Shannon's position- and momentum-space information entropies $S_{\rho}$ and $S_{\gamma}$ respectively. Both uncorrelated and correlated values of $S_{\rho}$ are decreasing function of $Z$ while those of $S_{\gamma}$ have been found to increase with the atomic number. But the relative contribution to $S_{\rho}$ due to inter-electronic repulsion increases as we go to high $Z$ atoms. But for $S_{\gamma}$ we observe the opposite. Recently, Lin and Ho \cite{22} made use of a Hylleraas-type wave function involving more than 400 parameters to obtain numbers for Shannon position-space information entropies by using a purely numerical routine. The results obtained by them for $H^-$, $He$ and $Li^+$, although somewhat augmented to our results, support our viewpoint for the effect of correlation on $S_{\rho 0}$. But the work in ref.22 could not be extended to deal with the corresponding momentum-space problem.
\par We attribute the observed variation of $S_{\rho}$ or $S_{\gamma}$ with atomic number to the interplay between interactions in the core and valence regions of the atom and thus provide a geometrical realization for the variation. More significantly, it is shown that correlation plays a more dominant role in modifying the bare information entropies than it does in correcting the entropy sum as well as Hartree-Fock binding energies of atomic systems. In view of the observed overall exaggerated response of Shannon information entropies to inter-electronic repulsion we suggest that either $S_{\rho}$ or $S_{\gamma}$ could be used as a correlation measure. In the literature there are other  useful suggestions for electron correlation tools. For example, Romera and Dehesa \cite{23} introduced a measure of correlation in two-electron systems by analyzing the product of Shannon entropy power and the Fisher information of the electron density. Almost simultaneously with this work, Huang and Kais \cite{24} proposed entanglement as an alternative measure of electron correlation. Since entanglement is one of the most striking phenomena of quantum mechanics and, at the same time, is directly observable, the work in ref. 24 deserve some special attention.
\par A few years ago, some serious attempts were made to construct accurate but simple correlated wave functions for a large number of two-electron systems \cite{25, 26}. There also exist simple correlated wave functions for the K- shell electrons of many neutral atoms \cite{27}. The wave functions in these works differ among themselves only in their dependence on the inter-electronic separation. The analytical model developed in this paper can easily be extended to work with each of these two-electron wave functions. It will, therefore, be interesting to apply our analytical model to systems considered in refs. 13 and 25 to 27 in order to provide further evidences for exaggerated response of Shannon's information entropies  to inter-electronic repulsion. 

\end{document}